\documentclass[11pt,twoside]{article}

\usepackage{asp2006}
\usepackage{epsf}
\usepackage{lscape}

\begin{document}
\title{MASYS -- The \textit{AKARI} spectroscopic survey of \\Symbiotic Stars in the Magellanic Clouds}   

\author{Rodolfo Angeloni$^{1,2}$, Stefano Ciroi$^1$, Paola Marigo$^1$, Marcella Contini$^{2,1}$, Francesco Di Mille$^{1}$ and Piero Rafanelli$^1$}   
\affil{
$^1$ Department of Astronomy, University of Padova, Italy \\
$^2$ School of Physics \& Astronomy, Tel Aviv University, Israel
}    

\begin{abstract} 
MASYS is the \textit{AKARI} spectroscopic survey of Symbiotic Stars in the Magellanic Clouds, and one of the European Open Time Observing Programmes approved for the \textit{AKARI} (Post-Helium) Phase-3. It is providing the first ever near-IR spectra of extragalactic symbiotic stars. The observations are scheduled to be completed in July 2009.
\end{abstract}



\section{Introduction}  
Symbiotic stars (hereafter SSs) are long-period interacting binaries composed of a hot compact star - generally but not necessarily a white dwarf - and an evolved giant star, whose mutual interaction via accretion processes is at the origin of extended emission recorded from radio to X-rays. 
Nowadays, 173 Galactic SSs are known, plus 26 suspected ones (Belczynski et al. 2000). Nonetheless, this value is in striking contrast with the \textit{predicted} total number of SSs in the Galaxy that, according to different estimates, oscillates between $3\times10^3$ (Allen 1984) and $4\times10^5$ (Magrini et al. 2003). The actual consistency of symbiotic population is thus a key-point to be investigated and has recently triggered specific observational campaigns (Corradi et al. 2008).\\
Whatever the case, SSs represent unique laboratories for studying a variety of important astrophysical problems and their reciprocal influence: e.g. nova-like thermonuclear outbursts (Munari et al. 1997), formation and collimation of jets (Tomov 2003), PNe morphology (Corradi 2003), and variable X-ray emission (Mukai et al. 2007). As binary systems, they offer a powerful benchmark to study the effect of binary evolution on the nucleosynthesis, mixing and dust mineralogy which characterize the AGB companion, likely different from what expected in single AGB stars (Marigo et al. 2007, 2008); moreover, they have even been proposed as potential progenitors of Supernovae \textit{Ia} (Munari et al. 1992; Hachisu et al. 1999, Lv et al. 2009).\\

The energetics operating these binaries is a basic ingredient for a meaningful understanding of the physical processes at work in SSs, but it relies on accurate knowledge of the distance. Unfortunately, the distances to Galactic SSs are largely uncertain, preventing reliable calibration of absolute stellar luminosities. As a matter of fact, such luminosities are of vital importance for evaluating the evolutionary status of SSs and the energy budget involved in the observed outbursts. 

\section{Extragalactic Symbiotic Stars} 
The specific stellar nature of symbiotic systems places them amongst the intrinsically brightest variable stars, easily detectable in nearby galaxies, in particular in the Magellanic Clouds (hereafter MCs). At present, 17 extragalactic SSs are known, 14 of which belonging to the MCs: 6 in the Small and 8 in the Large MC. It is clear that observations of Magellanic SSs remove the primary ambiguity in studying these interacting stars, namely the lack of reliable distances. It has even been suggested to use them as standard candles for the calibration of Galactic objects (Vogel et al. 1994), as already done with Novae.

Magellanic SSs (hereafter MSSs) are also interesting in themselves, because they differentiate from Galactic SSs in several aspects. SSs were classified into S- and D-types according to whether the cool star (S-type) or dust (D-type) dominated the near-IR spectral range (Webster \& Allen 1975). In our Galaxy, D-type SSs host invariably a Mira variable, while RGB stars are generally found in S-types. As their Galactic cousins, MSSs contain low mass ($\leq$ 3 M$_{\odot}$) giants as cool components. However, in MSSs, only AGB stars are found (Kniazev et al. 2009). In LMC, 4 out of 8 SSs are classified as D-type systems (50\%, much higher than the Galactic ratio of $\sim$20\%), while no D-types have been found in SMC. These results are intriguing, because S-type Galactic SSs rarely contain AGB stars. Furthermore, the position in the H-R diagram of the hot component reveals that in MSS they are amongst the hottest and the brightest within the known symbiotic population (Mikolajewska 2004, Fig.2c). It is probable that some of these results stem directly from the nature of the present sample, strongly biased toward the brightest objects we have been able to detect so far. 

Nevertheless, it is remarkable that the lack of dusty SSs in SMC may somehow reflect the very low Z in SMC, and that the Galactic SS with the lowest measured Z$\sim$0.002 - AG Dra - is also amongst the hottest systems. Another point that makes MSSs interesting to study is that they offer a direct determination of chemical abundances in extragalactic giants, being the chemical composition of the cool component reflected in the emission line spectrum of the nebula (Nussbaumer et al. 1988). In the past years, this opportunity triggered several studies which, thanks to sparse X/UV/optical observations, tempted to constrain the nebular conditions and the chemical abundances of MSSs (Vogel et al. 1994, 1995; M$\ddot{u}$rset et al. 1996), leading to surprising yet unconfirmed results. 
For example, the nebular chemistry and dynamics as derived from the N and Ne emission lines suggest in some objects a recent mixing of stellar winds in the interbinary interaction zone that might be representative of a recent outburst activity; while the already mentioned lower Z in MCs with respect to Galactic values, along with the suggestion that in several MSSs silicon seems not to be depleted (Vogel et al. 1994), have been presented as an indication that none of these systems would have an appreciable amount of dust. Today, this interpretation appears too simplistic, as it has been shown that dust can actually form also at low metallicity (also in the same MCs, van Loon et al. 2008; Matsuura et al. 2009), and even in potentially adverse conditions like those found within strongly shocked environments (Williams 2008). 

The issue is that any metallicity effect on the symbiotic phenomenon is virtually unknown.\\

Therefore, it sounds as a kind of a paradox that the only wavelength range where a single spectrum may allow a direct investigations of nebular conditions, chemical abundances of stellar components, and dust grain properties such as temperature and composition - namely the Infrared - is also the one for which there is no data. The same classification between S- and D-type SSs, that by original definition was introduced accodingly to the SED profile in the range 1-5 $\mu m$, in the case of MSSs is established rather tentatively on the basis of a few optical emission line ratios, relying on the fact that a larger binary separation (as in D-types) should turn into a lower electron density.

\section{\textit{MASYS}} 
The \textit{AKARI} (\citet{murakami07}) spectroscopic survey of \textbf{MA}gellanic \textbf{SY}mbiotic \textbf{S}tars (MASYS -- Phase-3 approved proposal in the European Time -- PI: Angeloni) is observing the whole sample of known MSSs (Table \ref{tab:lit}) in both the moderate (NP) and high resolution (NG) spectroscopic mode of IRC (\citet{onaka07}). We emphasize that these observations will provide \emph{the first ever NIR spectra of Symbiotic Stars in the Magellanic Clouds}, giving a fundamental and unique contribution to the understanding of the symbiotic phenomenon in its diverse appearance and, in general, of dust properties in complex environments at different metallicity. Bearing in mind the diverse morphology of symbiotic spectra in this wavelength region, at the present time  only hypothesis can be made about the detectable spectral features we are going to detect. Hopefully, these data would allow, among others, 1) to look for photospheric molecular bands (e.g. OH, HCl, H$_2$O, CO), that can inform about the nature of the cool component, so far mainly inferred by means of optical spectra; 2) look for emission lines (e.g. [SiVII]2.48$\mu$m, [FeVII]2.62$\mu$m, HBr$\gamma$ 2.626$\mu$m, [MgIV]4.49$\mu$m, [ArVI]4.52$\mu$m) which help constrain the physical conditions in the photoionized, shocked nebula (as done for Galactic SSs in Angeloni et al. 2007a); 3) look for specific dust bands, whose strengths and profiles are capable of revealing the chemical composition of dust grains and the physical conditions of the dust condensation environment (Angeloni et al. 2007a,b). \\

\textit{MASYS} observations are scheduled to be completed in July 2009.

\begin{table*}
\caption{\footnotesize Magellanic symbiotic stars in the literature. \label{tab:lit}}
\begin{footnotesize}
\begin{center}
\begin{tabular}{c c c c c c}\\
\hline \hline
 Symbiotic name$^a$ & Coordinates & Type$^a$ & K mag.$^b$ & Integrated flux \\
 & $\alpha \; [J2000] \;\delta$ &  &  & in the K band [mJy] \\
\hline
 SMC 1 & 00:29:10.800 $\;$ -74:57:39.91 & S & 12.64 & 5.86 \\
 SMC 2 & 00:42:47.975 $\;$ -74:41:59.88 & S & 13.10 & 3.83\\
 SMC 3 & 00:48:20.05 $\;$ -73:31:52.2 & S & 10.80 & 31.9 \\
 SMC-N60 & 00:57:05.867 $\;$ -74:13:16.27 & ? & 12.62 & 5.99\\
 LIN 358 & 00:59:12.259 $\;$ -75:05:17.50 & S & 11.46 & 17.3\\
 SMC-N73 & 01:04:39.289 $\;$ -75:48:24.78 & S & 11.6$^a$ & 15.3 \\
 LMC-S154 & 04:51:50.3 $\;$ -75:03:36 & D & 10.1$^a$ & 60.8\\
 LMC-S147 & 04:54:04.6 $\;$ -70:59:34.0 & S & 11.9$^a$ & 11.6\\
 LMC-N19  & 05:03:24.0 $\;$ -67:56:35.0 & ? & - & - \\
 LMC 1 & 05:25:01.080 $\;$ -62:28:48.67 & D & 10.71 & 34.8 \\
 LMC-N67 & 05:36:07.653 $\;$ -64:43:22.46 & S & 11.4$^a$ & 18.04\\
 Sanduleak's star & 05:45:19.741 $\;$ -71:16:07.08 & D & 13.0$^a$ & 4.21\\
 LMC-S63 & 05:48:43.502 $\;$ -67:36:10.60 & S & 11.33 & 19.6\\
 SMP LMC 94 & 05:54:10.4 $\;$ -73:02:39 & ? & - & - \\
\hline
\end{tabular} 
\end{center} 
\flushleft
$^a$from Belczynski et al. 2000; \\
$^b$from 2MASS.\\
\end{footnotesize}
\end{table*}


\vspace*{\fill}

\acknowledgements 
MASYS is based on observations with AKARI, a JAXA project with the participation of ESA.
Rodolfo Angeloni would like to thank the \textit{AKARI Conference} SOC and LOC for the support received.


\end{document}